\documentclass[aps,preprint]{revtex4}
\usepackage{amsfonts}
\usepackage{amsmath}
\usepackage{amssymb}
\usepackage{graphicx}

\input{tcilatex}

\begin{document}

\title{Shocklets in the Comet Halley Plasma }
\author{Ismat Naeem$^{1,2}$, Zahida Ehsan$^{3,4,},$ A. M. Mirza$^{1}$ and G.
Murtaza$^{5}$}
\affiliation{$^{1}$Department of Physics, Quaid-e-Azam University (QAU), Islamabad
Pakistan}
\affiliation{$^{2}$Department of Physics, Faculty of Basic \& Applied Sciences,
International Islamic University, Islamabad, Pakistan}
\affiliation{$^{3}$Space and Plasma Astrophysics Research Group (SPAR), Department of
Physics, COMSATS University Islamabad, Lahore Campus 54000, Pakistan }
\affiliation{$^{4}$National Centre for Physics, Shahdara Valley Road, Islamabad 45320,
Pakistan. }
\affiliation{$^{5}$Salam Chair in Physics, G.C.University Lahore, 54000 Pakistan. }
\email{ehsan.zahida@gmail.com}
\date{\today}

\begin{abstract}
Dust acoustic (DA) waves evolving into shocklets are investigated in the
Comet Halley plasma system relaxing to Maxwellian, Kappa and Cairns
distributions. Here dynamics of dust is described by the fully nonlinear
continuity and momentum equations. A set of two characteristic wave
nonlinear equations is obtained and numerically solved to examine the DA
solitary pulse which develops into oscillatory shocklets with the course of
time such as at time $\tau=0,$ symmetric solitary pulses are formed, which
develop into oscillatory shocklets. It has been observed that variation in
superthermality strongly affects the profiles of nonlinear DA structures in
terms of negative potential, dust velocity and density.
\end{abstract}

\maketitle
\email{\"{y} }

\section{INTRODUCTION}

In our solar system, billions of comets can be found circumnavigating the
Sun. These are snowballs of frozen gases, rocks and cosmic dust particles.
It is widely believed that like nucleus in an atom, nucleus of the comet is
discrete and cohesive made of volatile solids containing all the mass of the
comet. While orbiting around the Sun, when a comet approaches closer to the
Sun, it expels dust particles and ionized gases in huge quantity into a
behemoth like brightening head bigger than most planets. This dust and gases
form a \ dusty plasma and is in the form of a tail that expands outwardly
away from the Sun for millions of miles. These dust grains are electrically
charged by virtue of plasma which permeates the solar system. From the
observation of dust impact analyzer installed on Vega spacecraft supported
that much of the carbon is trapped within dust grains \cite{1,2,3}.

Due to low geometric albedo ($\backsim0.04$), comets are considered darkest
in the solar system$.$ Whereas Halley's comet is the most luminous one which
appears after an interval of 76 years. Surface of the comet is covered with
a dust mantle and has temperature of $\backsim300-400$ K where the latter
was determined from the infrared spectrometer (IKS) installed on the \textit{%
Vega 2} spacecraft \cite{4}.

Whipple revitalize the discrete nucleus concept and developed a seminal
model \cite{5} which could potentially explain the dynamic affects on the
orbits of comets (such as nongravitational recoil effects) as well as other
vital features of cometary observations. Despite large volume ($\backsim500$
km$^{{\LARGE 3}})$, the small bulk density ($0.1\backsim0.2g$cm$^{{\LARGE -3}%
}$) can be due to the small mass of Halley's comet ($\backsim5\times 10^{%
{\LARGE 16}}-10^{{\LARGE 17}}g$) which has been estimated independent of
gravitational effects of its orbit \cite{6}.

H$_{{\Large 2}}$O is the main constituent of the volatile nucleus, usually
volatiles like CO would be caged within the water - ice lattice and form
clathrate hydrate. It is also anticipated that when water-ice sublimation
starts, significant coma activity initiates and this happens at about
distance of $2-3$ AU \cite{7} . In a new comet \ water ice is of amorphous
nature and transforms into crystalline structure as the kinetic energy of
the cometary nucleus increases and reaches to a critical value. At this
point usually an unanticipated nucleus burst out and this happens between
about 3 and 6 AU \cite{8} .

Mendis in 1986 calculated the flights of the dust grains of different sizes
(micron-submicron) that are anticipated to be expelled out of the cometary
nucleus. Observational data from \textit{Vega 1 \& 2} and \textit{Giotto}
spaceports shows that plasma of comae of comet Halley contains ions of both
positive and negative polarity other than cosmic dust and electrons. These
ions are such as hydroxides (OH$^{+}$, OH$^{-}),$hydrogen (H$^{+}$, H$^{-})$%
, oxygen (O$^{+}$, O$^{-})$, silicon (Si$^{+}$, Si$^{-}$) etc \cite{9, 10}.

This kind of plasma with additional negative ions known as \textit{%
electronegative} dusty plasma is chemically very reactive. Electronegative
plasma is also found in other astrophysical environments such as D-region of
the ionosphere, themesosphere, photosphere of Sun etc. and plays a vital
role in the semiconductor technology. In most of these astrophysical and
laboratory plasmas, electrons can stick to the surface of dust and so their
density reduces and such system can also be treated as electron depleted
plasma. Therefore, in this setting. role of negative ions becomes more
crucial \cite{10} .

Observations reveal excitation of foreshock and bow shock like nonlinear
structures at Halley's comet. Shock waves are quite common nonlinear
structures in astrophysical settings for instance are triggered when
interaction of solar wind with the comets or magnetic field of different
planets takes place, these waves are also anticipated to appear in galactic
dust clouds at the beginning stage of star creation. Waves\ of compressional
nature are generated when dust particles impact with solid surface of a
spacecraft, in both the dust grain and surface of the spacecraft.

Shocks are outcome of discontinuous disturbances traveling in a medium where
an abrupt jump in physical quantities like density, pressure and temperature
across some narrow region leads to such structures. Dissipation structures
like kinematic viscosity (also known as momentum diffusivity) feature of
collisions between dust and ions and the fluctuating charge of dust
particles lead to the formation of dust acoustic (dust ion) DA (DIA) shock
waves in dusty plasmas with $\Gamma_{c}<<1$ (coupling parameter). Whereas in
opposite case when plasma is strongly coupled ($\Gamma_{c}>>1$), shear and
bulk viscosities are considered crucial in the excitation of DA or DIA shock
waves of monotone and oscillatory nature \cite{11} . Trigger of shock waves
is an irreversible phenomena and\ contrary to nonlinear solitary waves,
there is a sharp dissipation of energy and speed with the distance of the
shock wave alone.

For the better understating of this important class of nonlinear waves in
dusty plasma which has enormous significance in the astrophysical context,
many experimental investigations are worth mentioning. For example first
observation of shock waves in radio frequency 3D dusty (complex) plasma
discharge under minute gravity condition was carried in the PKE-Nefedov
device by Samsonov et al. \cite{12} Latter a compressional DA shock wave
triggered with the aid of impulsive axial magnetic field was examined \cite%
{13}. In another interesting study, excitation of a bow shock in a 2D dusty
plasma was examined where a super-sonic flow of micro sized negatively dust
particles around a stationary object was used. this exactly mimics the
aforementioned situation when shock waves are generated due to the impact of
dust particles with the insitu spacecraft. Recently, Jaswal et al
experimentally studied the evolution of DA shock waves compared their
propagation characteristics numerical results based on
Korteweg-de-Vries-Burgers (KdV-Burger) equation \cite{11}.

There has been extensive theoretical investigation of DA and DIA shock waves%
\cite{14,15,16,17,18,19,20}, in one of earlier works by Ghosh, Ehsan, and
Murtaza \cite{11}, authors investigated shock waves deriving KdV-Burger
equation for the Comet Halley plasma where nonsteady dust charge variation
produces anomalous dissipation that leads to the shock wave. There authors
transformed KdV burger equation to another one analogous to damped
anharmonic oscillator and using fifth order Runge-Kutta technique,
oscillatory nature shock amplitudes were studied. It was reported that shock
wave is more dispersive in nature for a the higher strength of magnetic
field. whereas for the parallel propagation a monotonic shock was observed.

In the present manuscript we aim to present a fully nonlinear theory to
study large-amplitude DA waves developing into the DA shocklet dust acoustic
in the Cometary plasma using diagonalization method \cite{21} . For the
readers it is worth mentioning the term shocklet was introduced to represent
sporadic steep fronts observed in high-speed turbulent compressible fluid
and in space as flank formations associated with planetary-scale shocks \cite%
{19',19''} .

When solar wind passes through a comet makes it a complex plasma system
which is far away from the state of equilibrium. In this scenario, cometary
electrons and ions will asymptotically approach to a nonequilibrium steady
state with power-law distributions, e.g., often fitted with the Cairns,
Kappa or r, q distribution \cite{22,23,24} .

To the best of authors' knowledge this has not been studied earlier and
results of the present investigation are useful for understanding the
physics of formation of shocklets at Halley's comet.

The paper is organized in the following manner: In Sec. II, the basic set of
equations for the dust acoustic shock wave of comet Halley plasma are given.
Section III deals with the derivation nonlinear equation. In Sec. IV and V
provide quantitative analysis and conclusions, respectively.

\section{Basic Set of Equations}

Consider a one dimensional, collisionless unmagnetized plasma, whose
components are the four species consisting of electrons, singly ionized
negatively and positively charged ions with dust grains that are negatively
charged. We assumed that spherically shaped dust grains are carrying a
constant charge because the time for the dust acoustic shocklets \ to be
discussed is much smaller than that required for further significant change
in dust charge. The condition of charge neutrality at equilibrium gives $%
n_{i_{+}0}=n_{e0}-\varepsilon Z_{d0}n_{d0}+n_{i_{-}0},$ where $n_{p0}$
denotes the equilibrium number density of the $p$ th species (viz., $p=i_{+}$
for positively charged ions, $p=i_{-}$ for negatively charged ions, $p=e$
for electrons, $p=d$ for dust particles and $Z_{d0}$ is the equilibrium dust
charge state.\ The following set of one dimensional nonlinear fluid
equations are used, which describe the dynamics of DA solitary and shock
waves:%
\begin{equation}
\frac{\partial n_{d}}{\partial t}+v_{d}\frac{\partial n_{d}}{\partial x}%
+n_{d}\frac{\partial v_{d}}{\partial x}=0,   \label{1}
\end{equation}%
\begin{equation}
\frac{\partial v_{d}}{\partial t}+v_{d}\frac{\partial v_{d}}{\partial x}-%
\frac{Z_{d0}e}{m_{d}}\frac{\partial\phi}{\partial x}+\frac{3n_{d}v_{Td}^{2}}{%
n_{d0}^{2}}\frac{\partial n_{d}}{\partial x}=0   \label{2}
\end{equation}
and 
\begin{equation}
\frac{\varepsilon_{0}}{e}\frac{\partial^{2}\phi}{\partial x^{2}}%
=Z_{d0}n_{d}+n_{e}+n_{i_{-}}-n_{i_{+}},   \label{3}
\end{equation}
where $v_{Td}[=\left( k_{B}T_{d}/m_{d}\right) ^{1/2}]$, $v_{d}$ and $\phi$
represents the dust thermal speed, dust fluid velocity and the electrostatic
potential. As the singly charged negative, positive ions and electrons are
assumed to be in thermal equilibrium, so we can described their densities by
the following expressions:%
\begin{equation}
n_{s}\left( \phi\right) =n_{s0}\exp\left( \frac{e\phi}{k_{B}T_{s}}\right) , 
\label{4}
\end{equation}%
\begin{equation}
n_{i_{+}}\left( \phi\right) =n_{i_{+}0}\exp\left( -\frac{e\phi}{%
k_{B}T_{_{i_{+}}}}\right)   \label{5}
\end{equation}
for the case when electrons, singly charged negative and positive ions are
taken as nonthermal, we can expressed the number densities for Kappa
distributed electrons and singly charged negative and positive ions by the
following expressions \cite{21,23}:

\begin{equation}
n_{s}\left( \phi\right) =n_{s0}\left\{ 1-\left( \kappa-\frac{3}{2}\right)
^{-1}\frac{e\phi}{k_{B}T_{s}}\right\} ^{-\kappa+1/2},   \label{6}
\end{equation}%
\begin{equation}
n_{_{i_{+}}}\left( \phi\right) =n_{i_{+}0}\left\{ 1+\left( \kappa-\frac {3}{2%
}\right) ^{-1}\frac{e\phi}{k_{B}T_{_{i_{+}}}}\right\} ^{-\kappa+1/2} 
\label{7}
\end{equation}
and for the case of Cairns distributed electrons and singly charged negative
and positive ions densities are given by \cite{21, 24}

\begin{equation}
n_{s}\left( \phi\right) =n_{s0}\left\{ 1+\Gamma\left( -\frac{e\phi}{%
k_{B}T_{s}}\right) +\Gamma\left( -\frac{e\phi}{k_{B}T_{s}}\right)
^{2}\right\} \exp\left( \frac{e\phi}{k_{B}T_{s}}\right) ,   \label{8}
\end{equation}%
\begin{equation}
n_{_{i_{+}}}\left( \phi\right) =n_{i_{+}0}\left\{ 1+\Gamma\left( \frac{e\phi%
}{k_{B}T_{_{i_{+}}}}\right) +\Gamma\left( \frac{e\phi}{k_{B}T_{_{i_{+}}}}%
\right) ^{2}\right\} \exp\left( -\frac{e\phi}{k_{B}T_{_{i_{+}}}}\right) 
\label{9}
\end{equation}
where the subscript $s$ equals $e$ for electrons and $i_{-}$ for negatively
charged ions. Here, $\Gamma=4\alpha/(1+3\alpha)$ comprise of Cairns
parameter $\alpha$ that determines the population of nonthermal particles.
It is important to mention here that if we consider $\kappa\rightarrow\infty$
and $\Gamma\rightarrow0$ in Eqs. $(6-9)$, respectively, then we can directly
replicate the Maxwell-Boltzmann density distribution for electrons,
positively and negatively charged ions.

\section{Nonlinear evolution equation}

In order to normalized the above set of Eqs. $(1-9)$ we apply the scaled
parameters \cite{21} as $N_{d}=n_{d}/n_{do}$, $V_{d}=v_{d}/c_{d\text{ }%
},U=e\phi/k_{B}T_{e},$ $\xi=x/\lambda_{0}$ and $\tau=t\omega_{pd}$ where $%
C_{da}[=\omega_{pd}\lambda_{0}=$ $\left( Z_{d0}k_{B}T_{e}/m_{d}\right)
^{1/2}]$ represents the DA speed with $\omega_{pd}=\left(
Z_{d0}^{2}e^{2}n_{do}/\varepsilon_{0}m_{d}\right) ^{\frac{1}{2}}$the dust
plasma frequency and $\lambda_{0}=\left(
\varepsilon_{0}k_{B}T_{e}/Z_{d}e^{2}n_{do}\right) ^{\frac{1}{2}}$ is the
scale length. Hence, the normalized dust momentum and continuity equations
can be written as%
\begin{equation}
\frac{\partial V_{d}}{\partial\tau}+V_{d}\frac{\partial V_{d}}{\partial\xi }-%
\frac{\partial U}{\partial\xi}+3\frac{T_{d}N_{d}}{Z_{d0}T_{e}}\frac{\partial
N_{d}}{\partial\xi}=0,   \label{10}
\end{equation}
and%
\begin{equation}
\frac{\partial N_{d}}{\partial\tau}+V_{d}\frac{\partial N_{d}}{\partial\xi }%
+N_{d}\frac{\partial V_{d}}{\partial\xi}=0   \label{11}
\end{equation}
Mostly, the dust temperature effect is neglected because of low values of
dust temperature $T_{d}$ and also because of the large dust charge state so
we can neglect the last term in Eq. $(10)$ which gives the following form%
\begin{equation}
\frac{\partial V_{d}}{\partial\tau}+V_{d}\frac{\partial V_{d}}{\partial\xi }-%
\frac{\partial U}{\partial\xi}=0   \label{12}
\end{equation}
In the current work, we are intended to study nonstationary, one dimensional
nondispersive fully nonlinear DAW's. Thus, by neglecting the dispersive term
viz., $\partial_{x}^{2}\phi=0$ in Eq. $(3)$ and by utilizing Eqs. $(4)$ and $%
(5)$ for thermal distributed particles and Eqs. $(6-9)$ for the case of non
thermal distributed particles, yields the following results 
\begin{equation}
N_{d,b}(U)=\left\{ 
\begin{array}{c}
\lbrack-\delta_{+}\exp U+\exp\left( -\frac{U}{\sigma_{+}}\right) -\delta
_{-}\exp\left( \frac{U}{\sigma_{-}}\right) ]/\alpha_{d} \\ 
\text{for Maxwellian distributed ions and electrons } \\ 
\lbrack-\delta_{+}\left( 1-\frac{U}{\left( \kappa-\frac{3}{2}\right) }%
\right) ^{-\kappa+1/2}+\left( 1+\frac{U}{\sigma_{+}\left( \kappa-\frac {3}{2}%
\right) }\right) ^{-\kappa+1/2}-\delta_{-}\left( 1-\frac{U}{\sigma_{-}\left(
\kappa-\frac{3}{2}\right) }\right) ^{-\kappa+1/2}]/\alpha_{d} \\ 
\text{ for Kappa distributed distributed ions and electrons } \\ 
\lbrack-\delta_{+}\eta_{1}(U)\exp U+\eta_{2}(U)\exp(-\frac{U}{\sigma_{+}}%
)-\delta_{-}\eta_{3}(U)\exp(\frac{U}{\sigma_{-}})]/\alpha_{d} \\ 
\text{for Cairns distributed ions and electrons }%
\end{array}
\right.   \label{13}
\end{equation}
Here $\eta_{1}(U)=\left( 1-\Gamma U+\Gamma U^{2}\right) $, $\eta
_{2}(U)=\left( 1+\frac{\Gamma U}{\sigma_{+}}+\frac{\Gamma U^{2}}{\sigma
_{+}^{2}}\right) $ and $\eta_{3}(U)=\left( 1-\frac{\Gamma U}{\sigma_{-}}+%
\frac{\Gamma U^{2}}{\sigma_{-}^{2}}\right) .$The subscript $b$ $=$ $m$ for
Maxwellian distributed ions and electrons, $b=\kappa$ for Kappa distributed
non-Maxwellian negatively and positively charged ions and electrons and $b$ $%
=c$ for Cairns-distributed non-Maxwellian negatively and positively ions and
electrons. Where $\sigma_{+}(=T_{_{i_{+}}}/T_{e})$ represents the positive
ion to electron temperature ratio, $\sigma_{-}(=T_{_{i_{-}}}/T_{e})$
represents the negative ion to electron temperature ratio, $%
\delta_{+}(=n_{e0}/n_{i_{+}0})$ is the electron to positive ion density
ratio and $\delta_{-}(=n_{i-0}/n_{i_{+}0})$ is taken as negative to positive
ion density ratio which can be written in terms of dust concentration as $%
\alpha_{d}=1-\delta_{+}-\delta_{-}$. It is worth mentioning here that the
inequality $\delta_{+}+\delta_{-}<1$ always holds in order to have positive
dust concentration.

Substituting Eq. $(13)$ into Eqs. $(11)$ and $(12)$ and carrying out
differentiations w.r.t space-time coordinates $(\xi$ ,$\tau)$ leads to the
following form of equation 
\begin{equation}
\frac{\partial U}{\partial\tau}+V_{d}\frac{\partial U}{\partial\xi}-\chi
_{b}\left( U\right) \frac{\partial V_{d}}{\partial\xi}=0   \label{14}
\end{equation}
where 
\begin{equation}
\chi_{b}\left( U\right) =\left\{ 
\begin{array}{c}
\left. \exp\left( -\frac{U}{\sigma_{+}}\right) -\delta_{-}\exp\left( \frac{U%
}{\sigma_{-}}\right) -\delta_{+}\exp U/\delta_{+}\exp U+\frac {1}{\sigma_{+}}%
\exp\left( -\frac{U}{\sigma_{+}}\right) +\frac{\delta_{-}}{\sigma_{-}}%
\exp\left( \frac{U}{\sigma_{-}}\right) \right. \\ 
\text{ for Maxwellian distributed pair ions and electrons } \\ 
\left. -\delta_{+}\left( 1-\frac{U}{\left( \kappa-\frac{3}{2}\right) }%
\right) ^{-\kappa+1/2}+\left( 1+\frac{U}{\sigma_{+}\left( \kappa-\frac {3}{2}%
\right) }U\right) ^{-\kappa+1/2}-\delta_{-}\left( 1-\frac{U}{%
\sigma_{-}\left( \kappa-\frac{3}{2}\right) }U\right) ^{-\kappa +1/2}/\right.
\\ 
\left. \delta_{+}c_{k}\left( 1-\frac{U}{\left( \kappa-\frac{3}{2}\right) }%
\right) ^{-\kappa-1/2}+\frac{c_{k}}{\sigma_{+}}\left( 1+\frac{U}{\sigma
_{+}\left( \kappa-\frac{3}{2}\right) }\right) ^{-\kappa-1/2}+\frac {%
\delta_{-}}{\sigma_{-}}c_{k}\left( 1-\frac{U}{\sigma_{-}\left( \kappa -\frac{%
3}{2}\right) }\right) ^{-\kappa-1/2}\right. \\ 
\text{for Kappa distributed pair ions and electrons} \\ 
\left. \delta_{+}\eta_{1}(U)\exp U-\eta_{2}(U)\exp(-\frac{U}{\sigma_{+}}%
)+\delta_{-}\eta_{3}(U)\exp(\frac{U}{\sigma_{-}})/\right. \\ 
\left. -\delta_{+}\left[ \eta_{1}(U)+\zeta_{1}(U)\right] \exp U-\left[ \frac{%
\eta_{2}(U)}{\sigma_{+}}-\zeta_{2}(U)\right] \exp(-\frac{U}{\sigma_{+}})-%
\left[ \frac{\delta_{-}}{\sigma_{-}}\eta_{3}(U)+\delta_{-}\zeta _{3}(U)%
\right] \exp(\frac{U}{\sigma_{-}})\right. \\ 
\text{for Cairns distributed pair ions and electrons }%
\end{array}
\right.   \label{15}
\end{equation}
here $\zeta_{1}(U)=\left( -\Gamma+2\Gamma U\right) $, $\zeta_{2}(U)=\left( 
\frac{\Gamma}{\sigma_{+}}+\frac{2\Gamma U}{\sigma_{+}^{2}}\right) $, $%
\zeta_{3}(U)=\left( -\frac{\Gamma}{\sigma_{-}}+\frac{2\Gamma U}{\sigma
_{-}^{2}}\right) $ and$\ c_{k}=\left( 2\kappa-1\right) /\left(
2\kappa-3\right) $ is the parameter that represent hot electron
superthermality effects and for $\kappa\rightarrow\infty,c_{k}$ $\rightarrow1
$ displaying the reduction of Kappa distribution to standard Maxwellian
distribution. Equations $\left( 12\right) $ and $\left( 14\right) $ are the
normalized nonlinear coupled equations, which describes the large-amplitude
nonstationary DA shock waves.

We can express Eqs.$\left( 12\right) $ and $\left( 14\right) $ in the matrix
form:%
\begin{equation}
\frac{\partial}{\partial\tau}\left[ 
\begin{array}{c}
U \\ 
V_{d}%
\end{array}
\right] +\left[ 
\begin{array}{cc}
V_{d} & -\chi_{b}\left( U\right) \\ 
-1 & V_{d}%
\end{array}
\right] \frac{\partial}{\partial\xi}\left[ 
\begin{array}{c}
U \\ 
V_{d}%
\end{array}
\right] =0   \label{16}
\end{equation}
where the parameter $\chi_{b}\left( U\right) $ for each distributions is
expressed as given in Eq. ($15$)

The square matrix on the left hand side of Eq. $\left( 16\right) $ can be
diagonalized by means of a diagonalizing matrix technique. The two eigen
values of the square matrix can be resolved by det$\left( A-\lambda I\right)
=0$, given by%
\begin{equation}
\lambda_{\pm,b}=V_{d}\pm\sqrt{\chi_{b}\left( U\right) }   \label{17}
\end{equation}
where square matrix and unit matrix are given by

\begin{equation}
A=\left[ 
\begin{array}{cc}
V_{d} & -\chi_{b}\left( U\right) \\ 
-1 & V_{d}%
\end{array}
\right] \text{ and }I=\left[ 
\begin{array}{cc}
1 & 0 \\ 
0 & 1%
\end{array}
\right]   \label{18}
\end{equation}
The square matrix in Eq. $\left( 16\right) $ taken as $A$, can be
diagonalized by means of a diagonalizing matrix $C$ whose columns are the
eigenvectors of $A,$ so that%
\begin{equation}
C^{-1}AC=\left[ 
\begin{array}{cc}
\lambda_{+,b} & 0 \\ 
0 & \lambda_{-,b}%
\end{array}
\right]   \label{19}
\end{equation}
where 
\begin{equation}
C=\left[ 
\begin{array}{cc}
1 & 1 \\ 
-\sqrt{1/\chi_{b}\left( U\right) } & \sqrt{1/\chi_{b}\left( U\right) }%
\end{array}
\right]   \label{20}
\end{equation}
Multiplying Eq. $\left( 16\right) $ by $C^{-1}$from the left gives the
diagonalized system of equations%
\begin{equation}
\frac{\partial\Psi_{\pm}}{\partial\tau}+\lambda_{\pm,b}\frac{\partial\Psi
_{\pm}}{\partial\xi}=0   \label{21}
\end{equation}
Here the new variables are $\Psi_{\pm}=V_{d}\mp F\left( U\right) $ with $%
F\left( U\right) =\int_{0}^{U}\left[ \frac{1}{\chi_{b}\left( U\right) }%
\right] ^{\frac{1}{2}}dU$. A straightforward wave solution is found by
taking either $\Psi_{+}$or $\Psi_{-}$ to zero. Setting $\Psi_{-}$ to zero,
we obtain $V_{d}=-F\left( U\right) $ and $\Psi_{+}=2U_{b}$. Since $\Psi_{+}$%
can be written as function of $V_{d}$, so we can write Eq. $\left( 21\right) 
$ for $\Psi_{+}$ as 
\begin{equation}
\frac{\partial V_{d}}{\partial\tau}+\lambda_{+,b}\left( U\right) \frac{%
\partial V_{d}}{\partial\xi}=0   \label{22}
\end{equation}
where eigen value can be expressed in its new form as $\lambda_{+,b}\left(
U\right) =-F\left( U\right) +\sqrt{\chi_{b}\left( U\right) }$ where $%
\chi_{b}\left( U\right) $ for Maxwellian, Cairns and Kappa distributed
positively and negatively ions and electrons are given by Eq $\left(
15\right) $. Since we know that $V_{d}$ is a function of $U$ we can also
write an equation similar to Eq. $\left( 22\right) $ as%
\begin{equation}
\frac{\partial U}{\partial\tau}+\lambda_{+,b}\left( U\right) \frac{\partial U%
}{\partial\xi}=0   \label{23}
\end{equation}
The above equation interprets the self steepening of the negative potential $%
U$, the general solution of Eq. $\left( 23\right) $ is $U=U_{0}\left[
\xi-\lambda_{+,b}\left( U\right) \tau\right] $, where $U_{0}$ is a function
of one variable and is resolved by the initial condition for $U$ at $\tau=0$%
, where $\lambda_{+,b}\left( U\right) $ is the effective phase speed which
is a function of $U$, makes the general solution nonlinear and may therefore
self-steepen which develops into oscillatory shocks with the course of time
as depicted from the numerical plots shown in Fig. $1$. The speeds with
which these kind of shock fronts propagate is stated by Rankine Hugoniot
condition $v_{shock}=\{\Lambda_{L}\left( U_{L}\right) -\Lambda_{R}\left(
U_{R}\right) \}/\left( U_{L}-U_{R}\right) $, where $U_{R}\left( U_{L}\right) 
$ represents the value of $U$ on the right side (on the left side) of shock
front, and $\Lambda_{L}\left( U_{L}\right) $ and $\Lambda _{R}\left(
U_{R}\right) $ gives the left and right flux functions, where the flow
function is defined as $\Lambda\left( U\right) =\int_{0}^{U}\lambda
_{+}\left( U\right) dU$.

\section{Quantitative Analysis}

In this section, we solve numerically Eq. $\left( 23\right) $ for exploring
the temporal evolution of large amplitude localized DA waves in Comet Halley
plasma with Maxwellian and non Maxwellian like Kappa and Cairns distributed
ions and electrons. For illustration we have chosen some typical numerical
parameters of the Comet Halley taken at $\sim10^{4}$ km from the nucleus 
\cite{10, 27} such as are $n_{d0}\sim10^{5}$m$^{-3}$, positive ion density $%
n_{i_{+}0}$ $\sim2\times10^{8}$m$^{-3}$, $m_{+i}=m_{-i}\sim1.67\times 10^{-27%
\text{ }}$kg, $T_{e}=T_{i+}=T_{i-}=100$ eV, mass density for ice dust grains 
$\rho_{d}\sim9\times$ $10^{2}$kg/m$^{-3}$, $Z_{d0}=10^{5}$\ dust\ grains
with diameter $r_{0}\approx5\mu m$\ and we used equilibrium charge
neutrality condition to obtain the negative ions to positive ions density
ratio\ $\delta_{-}\left( =n_{i_{-}0}/n_{i_{+}0}\right) $. For these
parameters, we have dust plasma frequency $\omega_{pd}=1.79\times10^{-5}$Hz,
dust fluid velocity $C_{d}=1.30\times10^{-5}$ m/s, characteristic lengths $%
C_{d}/\omega_{pd}=0.743$ m and initial condition for \ electrostatic
potential is taken as $U=U_{0}\func{sech}\left[ \xi/d\right] $ with
amplitude $U_{0}=-0.15$ and pulse width $d=2.3.$

Figures 1 represents the time enhancement of normalized (a) electrostatic
negative potential (b) dust velocity (c) dust number density as a function
of position $\xi(=x/\lambda_{0})$ for dotted, solid and dashed curves
corresponding to different values of parameters $\kappa(=2,3,10)$ \ for
Kappa distribution in the right section [Figs. 1(a)-(c)] and for Cairns
distributed ions and electrons by increasing the parameters $%
\alpha(=0.1,0.3,0.5)$ for dotted, solid and dashed curves that determines
the population of the non thermal ions and electrons are shown in the right
panel [Figs. 1(d)-(f)] with fixed electron to positive ion number density
ratio $\delta_{+}=$ $0.984$, negative to positive ion number density ratio $%
\delta_{-}=0.011$, positive ion to electron temperature ratio $\sigma_{+}=1$
and negative ion to electron temperature ratio $\sigma_{-}=1$. It is
displayed that the negative potential pulses overlap on one another at time $%
\tau=0$, without showing any impact of superthermality. Although, the pulses
symmetry break with the passage of time $\tau=(3,4.5,6)$ s and consequently,
solitary pulses transform into an oscillatory shocks with enhanced self
steepness and wave amplitude. The similar trend for potential profile is
obtained for Cairns distributed ions.

Variation of ion and electron superthermality in Kappa distribution results
in the reduction of amplitude of solitary pulses and shocklets this results
for the reason that the effective phase speed $\lambda_{+,b}\left( U\right) $
become larger at a smaller value of $\kappa$ and vice versa. The effect of
increasing pair ions and electrons superthermailty with the time evolution
of non linear structures associated with the dust fluid velocity and density
profiles reduces the wave amplitudes and pulse width which are better
apparent as compare to the profiles of negative potential. On the contrary
in comparison with Kappa distribution, the Cairns distributed non thermal
pair ions and electrons with the time evolution of non linear structures
related with the negative potential, dust fluid velocity and density
profiles leads to the reduction of wave self steepness, amplitude and pulse
width of solitary waves and shocklets with increasing the parameter $%
\alpha(=0.1,0.3,0.5)$. These results may be explained by looking at the
particle distribution functions curves of Kappa and Cairns distribution.
Kappa distribution at high energies represent power law tails, however
Cairns distribution displays strong variations in shoulders as compared with
the tails so Kappa distribution is the most energetic distribution, whereas
Cairns distribution is least energetic which outcomes in the reduction of
linear and nonlinear effective phase speed with reduced self steepness in
comparison with Kappa distribution.

Figure 2 displays temporal development of nonlinear DA waves related with
the normalized (a) negative potential, (b) dust velocity and (c) dust
density profiles as a function of $\xi$ position for Maxwellian distribution
with fixed $\delta_{+}=$ $0.984,$ $\delta_{-}=0.011$ and $%
\sigma_{+}=\sigma_{-}=1$. It is shown that as time progresses the solitary
pulse stimulations propagate with non linear phase speed over the right hand
side of the plot and evolve into shocks with increase wave amplitude and
self steepness. It is important to mention here that DA excitations for the
glimpses of potential, dust fluid velocity and dust density in case of Kappa
distributed non thermal pair ions and electrons for high value of $\kappa=50$
and for Cairns distributed non thermal pair ions and electrons at $\alpha=0$
almost coincides with the case of Maxwellian as depicted for Kappa
distribution in the left panel [Figs. 3 (a)-(c)] and for Cairns distribution
in the right panel [Figs. 3 (d)-(f)].

Figure 4 and 5 \ manifests the variation of normalized dust number density
with ratio of electron to negative ion number density $\beta=n_{e0}/n_{i-0}$
[Fig. 4] and with electron to negative ion temperature ratio $T=T_{e}/T_{i-}$%
[Fig. 5] for Maxwellian (blue curve) Kappa (red curve) and Cairns (black
curve) distributed ions and electrons respectively with fixed $\delta
_{-}=0.011$, $\kappa=2$, $\alpha=0.1$ and $\sigma_{+}=\sigma_{-}=1$. It is
determined that increasing value of $\beta$ results in an enhancement of
dust number density $N_{d}$. On the other hand, by increasing the electron
to negative ion temperature ratio $\xi$ means an increase in the dust number
density.

It is evident from the curves of Figures 4 and 5 that dust number density is
higher for Kappa distribution and least for Cairns distribution and
intermediate for Maxwellian distribution.

\section{Conclusions}

In this paper we have studied the formation of large-amplitude dust acoustic
waves and their evolution into shocklets with time for a an unmagnetized
Maxwellian and non-Maxwellian Kappa/Cairns distributed comet Halley plasma
comprising of electrons, negative and positive ions in addition to
negatively charged dynamical dust grains. We worked out the fully nonlinear
hydrodynamic equations, momentum and continuity equations alongwith a
quasineutrality equation for the dust fluid by employing diagonalization
method technique, a set of two characteristic wave equations are found which
are then solved numerically. Our numerical results reveals the existence of
oscillatory shocks in a Maxwellian and non-Maxwellian dusty plasma with the
course of time. The effects of ion superthermality $\kappa$ and $\alpha$
parameter density are investigated on the profiles of solitary and
oscillatory shock waves resulting considerable variations in the wave
amplitudes and widths with the temporal development. The effect of electron
to negative ion temperature ratio $\xi$ and electron to negative ion density
ratio $\beta$ on dust number density are also investigated and found that it
is higher in case of Kappa distributed negative and positive ions and
electrons lower for Cairns distribution and intermediate for Maxwellian
distribution. It was found that the drift\ solitary and shock wave amplitude
is lower for cairns distributed pair ions and electrons as compare to Kappa
and Maxwellian distributions. To the best of author's knowledge present
study has not done before and is important for the comet Halley plasmsa.

\textbf{Acknowledgement:} One of us (Z.E) is grateful to the office of CAAD,
National Center for Physics (NCP)\ for the hospitality.

\textbf{Figure Captions}

Figure 1 (color online): Temporal change of the normalized (a)\ negative
potential $U\left( =e\phi/k_{B}T_{e}\right) $, (b) dust fluid velocity $%
V_{d}=v_{d}/c_{d\text{ }}$, and (c) dust density $N_{d}\left(
=n_{d}/n_{d0}\right) $ is plotted against the normalized position $\xi\left(
=x/\lambda_{0}\right) $ by taking different values of $\kappa=2$ (dotted
curve), $\kappa=$ $3$ (solid curve), and $\kappa=$ $10$ (dashed curve) in
the left section and $\alpha=0.1$ (dotted curve), $\alpha=$ $0.3$ (solid
curve), and $\alpha=$ $0.5$ (dotted curve) in the right section with fixed
values of $\delta_{+}=$ $0.984$, $\delta_{-}=0.011$ and $\sigma_{+}=%
\sigma_{-}=1$.

Figure 2 (color online): Temporal change of the normalized (a)\ negative
potential $U\left( =e\phi/k_{B}T_{e}\right) $, (b) dust velocity $%
V_{d}=v_{d}/c_{d\text{ }}$, and (c) dust density $N_{d}\left(
=n_{d}/n_{d0}\right) $ is plotted against the normalized position $\xi\left(
=x/\lambda_{0}\right) $ for Maxwellian distributed pair ions and electrons
with fixed values of $\delta_{+}=$ $0.984$, $\delta_{-}=0.011$ and $\sigma
_{+}=\sigma_{-}=1$.

Figure 3 (color online): Temporal change of the normalized (a)\ negative
potential $U\left( =e\phi/k_{B}T_{e}\right) $, (b) dust velocity $%
V_{d}=v_{d}/c_{d\text{ }}$, and (c) dust density $N_{d}\left(
=n_{d}/n_{d0}\right) $ is plotted against the normalized position $\xi\left(
=x/\lambda_{0}\right) $ for Kappa distribution with $\kappa=50$ is plotted
in the left section and for Cairns distribution with $\alpha=0$ is plotted
in the right section with fixed values of $\delta_{+}=$ $0.984$, $%
\delta_{-}=0.011$ and $\sigma_{+}=\sigma_{-}=1$.

Figure 4 (color online): The normalized dust density $N_{d}\left(
=n_{d}/n_{d0}\right) $ is plotted against the electron to negative ion
number density $\beta=n_{e0}/n_{i-0}$ for Maxwellian (blue curve) Kappa (red
curve) and Cairns (black curve) distributed pair ions and electrons
respectively with fixed $\kappa=2$, $\alpha=0.1$, $\delta_{-}=0.011$ and $%
\sigma_{+}=\sigma _{-}=1\mathrm{.}$

Figure 5 (color online): The normalized dust density $N_{d}\left(
=n_{d}/n_{d0}\right) $ is plotted against the electron to negative ion
temperature ratio $T=T_{e}/T_{i-}$ for Maxwellian (blue curve), Kappa (red
curve) and Cairns (black curve) distributed pair ions and electrons
respectively with fixed $\kappa=2$, $\alpha=0.1$, $\delta_{-}=0.011$ and $%
\sigma_{+}=\sigma_{-}=1\mathrm{.}$

\end{document}